# Influences of the Minkowski–Bouligand Dimension on Graphene-Based Quantum Hall Array Designs

D. S. Scaletta,[1] N. T. M. Tran,[2,3] M. Musso,[4] V. Ortiz Jimenez,[2] H. M. Hill,[2] D. G. Jarrett,[2] M. Ortolano,[4] C. A. Richter,[2] D. B. Newell,[2] and A. F. Rigosi[2,a]

[1]Department of Physics, Mount San Jacinto College, Menifee, California 92584, USA

[2]Physical Measurement Laboratory, National Institute of Standards and Technology (NIST), Gaithersburg, Maryland 20899, USA

[3]Joint Quantum Institute, University of Maryland, College Park, Maryland 20742, USA

[4]Department of Electronics and Telecommunications, Politecnico di Torino, Torino 10129, Italy

[a] Author to whom correspondence should be addressed.  email: afr1@nist.gov

This work elaborates on how one may develop high-resistance quantized Hall array resistance standards (QHARS) by using star-mesh transformations for element count minimization. Refinements are made on a recently developed mathematical framework optimizing QHARS device designs based on full, symmetric recursion by reconciling approximate device values with exact *effective* quantized resistances found by simulation and measurement. Furthermore, this work explores the concept of fractal dimension, clarifying the benefits of both full and partial recursions in QHARS devices. Three distinct partial recursion cases are visited for a near-1 GΩ QHARS device. These partial recursions, analyzed in the context of their fractal dimensions, offer increased flexibility in accessing desired resistance values within a specific neighborhood compared to full recursion methods, though at the cost of the number of required devices.

Graphene, when grown epitaxially (EG), has been developed into devices for electrical metrology due to its robust quantum Hall effect [1-3], and most EG-based devices that are used as resistance standards operate at the resistance plateau formed by the $v = 2$ Landau level ($R_H = \frac{1}{2}\frac{h}{e^2} \approx 12906.4037$ Ω). This single-value constraint significantly limits the equipment and general measurement infrastructure with which one may disseminate the unit of the ohm. Two dominant approaches to remove these limitations include: (1) quantum Hall array resistance standards (QHARS), which connect multiple Hall elements in series or parallel, and (2) the use of *p-n* junctions, where both approaches yield resistances of $qR_H$ (*q* being a positive rational number) [4-17].

Obtaining a large total area of high-quality EG, currently limited to the centimeter scale [18], remains a crucial step for designing any QHARS device requiring many individual Hall bar elements. With the lateral dimensional constraints, there is an inherent limitation on the total number of attainable QHARS elements through fabrication. As an example, an array that has 500 elements in series yields a maximum quantized resistance of approximately 6.5 MΩ, and this order of magnitude is much smaller than the range of resistances currently calibrated globally – up to PΩ levels in some cases [19-21]. Future QHARS devices may use star-mesh transformations that can achieve resistances at the highest levels of necessity [22-27].

This work explores a recently established, mathematical framework seeking to minimize the required number of elements in a QHARS to achieve high effective quantized resistances [22]. One of the key points of the framework was to optimize a star-mesh QHARS device design using full and symmetric recursion principles, and though this approach allows one to rapidly calculate an optimal device design, the designed device itself may hold a slightly different value than desired because of approximations inherent to the initial framework. Those discrepancies are understood and explained herein. Presented data from an approximately 1 GΩ QHARS device also support the underlying principles of this work.

In addition to the framework adjustments to full recursions, which are one form of pseudofractal in the context of the topology of the QHARS device design, other example pseudofractals are explored to reveal the benefits of partial recursions in QHARS device designs. All pseudofractals are examined via their Minkowski–Bouligand Dimension (MBD) [26], a measure that allows one to quantify complexity and correlate it with the flexibility of a QHARS device design when subjected to minor modifications. The analysis presented here offers strong advantages to a metrologist because such information may help prevent a full QHARS device from being rendered unusable due to a single grounded Hall element failure. All analyses are applicable to material systems that exhibit the quantum Hall effect, as well as artifact standard resistors.

QHARS devices were fabricated from square SiC chips measuring 7.6 mm × 7.6 mm that underwent a silicon sublimation procedure described in Ref. [24, 28]. Three main steps summarize device development: (1) growth, (2) fabrication, (3) post-fabrication and packaging. Films of EG were grown in a furnace and inspected using optical and confocal laser scanning microscopy [29], followed by fabrication of device contacts composed of NbTiN [18, 30]. QHARS devices were then measured in a cryostat at approximately 2 K with a Dual Source Bridge (DSB) [24].

Recall some of the fundamental principles and conclusions of Ref. [22], where the mathematical relationship between a star network and its equivalent mesh network ($N$ is equal in both networks, but the mesh contains one fewer node) is:

$$q_{ij} = q_i q_j \sum_{\alpha=i}^{N} \frac{1}{q_\alpha}$$

(1)

In Eq. 1, the indices go as high as $N$ with the condition that $i \neq j$ (and the definition of $q \equiv \frac{R}{R_H}$, where $q$ is defined as the number of single Hall elements exhibiting $\nu = 2$ quantization). Note that $q$, also known as the *coefficient of effective resistance*, is restricted to the set of positive integers ($q: q \in \mathbb{Z}^+$). One may also recall the parameter $M$, or recursion number, as well as the intended number of cleanroom-fabricated elements (represented by $q_{M:i}$ (single index) [22]), and the *effective number of elements* (represented by $q_{M:ij}^{(approx)}$ (two indices)). With $\xi$ as the number of grounded branches, the total number of elements in the final QHARS device is:

$$D_T(M, \xi, q_{M:ij}) = \frac{2^M}{\xi}\left(\xi q_{M:ij}^{(approx)} + 1\right)^{2^{-M}} - \frac{2^M}{\xi} + (2^M - 1)\xi$$

(2)

This function of three variables was used as a starting point for device design optimization. In Ref. [22], it was found that for a *desired* quantized resistance ($q_{M:ij}^{(approx)}$), a global minimum in Eq. 2 would correspond to values of $M$ and $\xi$, typically as non-integers. The next step would be to round $M$ and $\xi$ to the nearest integer and calculate $q_{M:i}$ in Eq. 3, followed by rounding of $q_{M:i}$ to the nearest integer. Now with all three integers determined, one may find the *effective* (near-exact) number of elements represented by $q_{M:ij}^{(near-exact)}$ (two indices):

$$q_{M:ij}^{(near-exac\,)} = \frac{1}{\xi}(\xi q_{M:i} + 1)^{2^M} - \frac{1}{\xi}$$

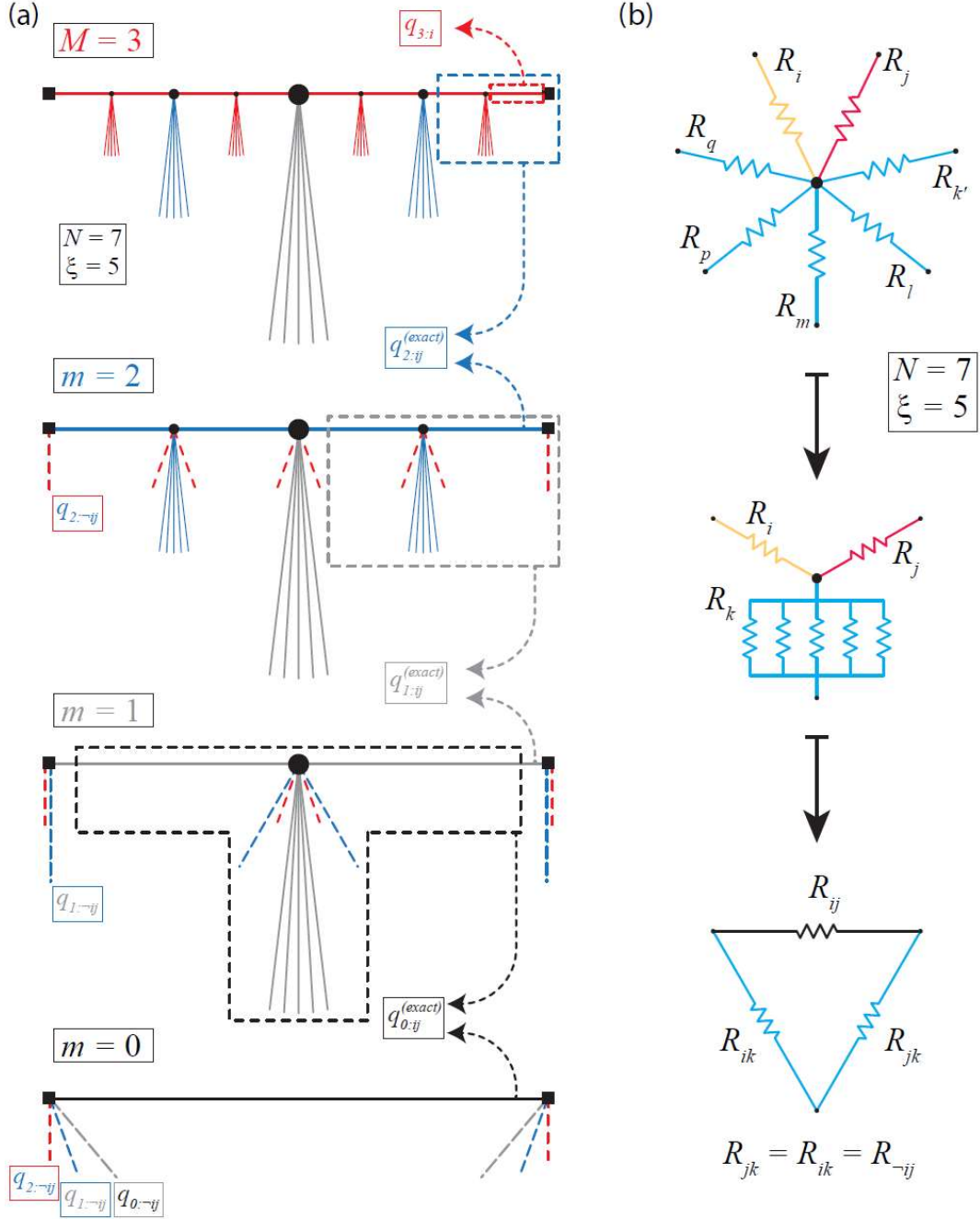

FIG. 1. (a) This illustration serves to guide one's intuition on how some virtual resistances ($q_{m:\neg ij}$) contribute to the overall resistance network (with an example focused on a device design where $M = 3$). After the full optimization process, the quantity $q_{M:i}$ always remains an exact number of elements, whereas every other form of $q$ represents an *effective* resistance. With the inclusion of virtual resistances from every star-mesh transformation, calculating $q_{m<M:ij}^{(exact)}$ becomes more cluttered. (b) One approach to simplifying the manual procedure of recalculating $q_{0:ij}^{(exact)}$, which is the quantity one should expect to measure from the device, is to treat all $\xi$ and $q_{m:\neg ij}$ branches as parallel branches in a single Y-Δ network.

When the *effective* (near-exact) number of elements was calculated in Ref. [22], it provided a more accurate numerical value for the expected quantized resistance output of the QHARS with the corresponding design parameters $M$, $\xi$, and $q_{M:i}$ (all integers). The issue is that this optimization process, centered on minimizing element count, contained some minor necessary approximations, without which one would not have the straightforward ability to calculate a device design within the neighborhood of a desired value of quantized resistance. The minor approximation, in short, is the exclusion of virtual resistors that contribute to the overall network when utilizing the star-mesh transformation. Examples of how these virtual resistances contribute are shown in Figure 1. One may argue that the optimization accomplished its objective of providing the proper neighborhood of quantized resistance in an efficient manner and that the analog electronic circuit simulator LTspice can provide the exact resistance expected [16, 31-33]; however, it would be appropriate to summarize how the exactness may be recovered mathematically.

The illustration in Fig. 1 (a) shows that, for an example final $M$ = 3 device design, some virtual resistances ($q_{m:\neg ij}$) contribute to the overall resistance network. Here, one defines $q_{m:\neg ij}$ as the virtual (and still technically *effective*) resistance branches that emerge after a star-mesh transformation is applied (the symbol $\neg ij$, or "not *ij*", indicates any other *effective* resistance that does contribute to the dominant *effective* resistance that is desired – see the bottom of Fig. 1 (b)). After the full device design optimization process, the quantity $q_{M:i}$ always reflects an exact number of elements, whereas every other form of $q$ represents an *effective* resistance.

With every star-mesh transformation contributing virtual resistances, calculating $q_{m<M:ij}^{(exact)}$ becomes more cumbersome, and one way to simplify the manual procedure of recalculating $q_{0:ij}^{(exact)}$, which is the quantity one should expect to measure from the device, is to treat all $\xi$ and $q_{m:\neg ij}$ branches as parallel branches in a single Y-Δ network (as in Fig. 1 (b)). These conditions allow one to formulate the mathematical corrections as:

$$q_{m:ij}^{(exact)} = 2q_{m+1:ij}^{(exact)} + \xi\left(q_{m+1:ij}^{(exact)}\right)^2 + 2\xi\frac{\left(q_{m+1:ij}^{(exact)}\right)^2}{q_{m+1:\neg ij}}$$

(4)

It is important to note that the last term of Eq. 4 vanishes for $q_{m=M:ij}^{(exact)}$ since it does not represent a virtual (*effective*) resistance. Furthermore, the definition of $q_{m:\neg ij}$ becomes straightforward from Eq. 1:

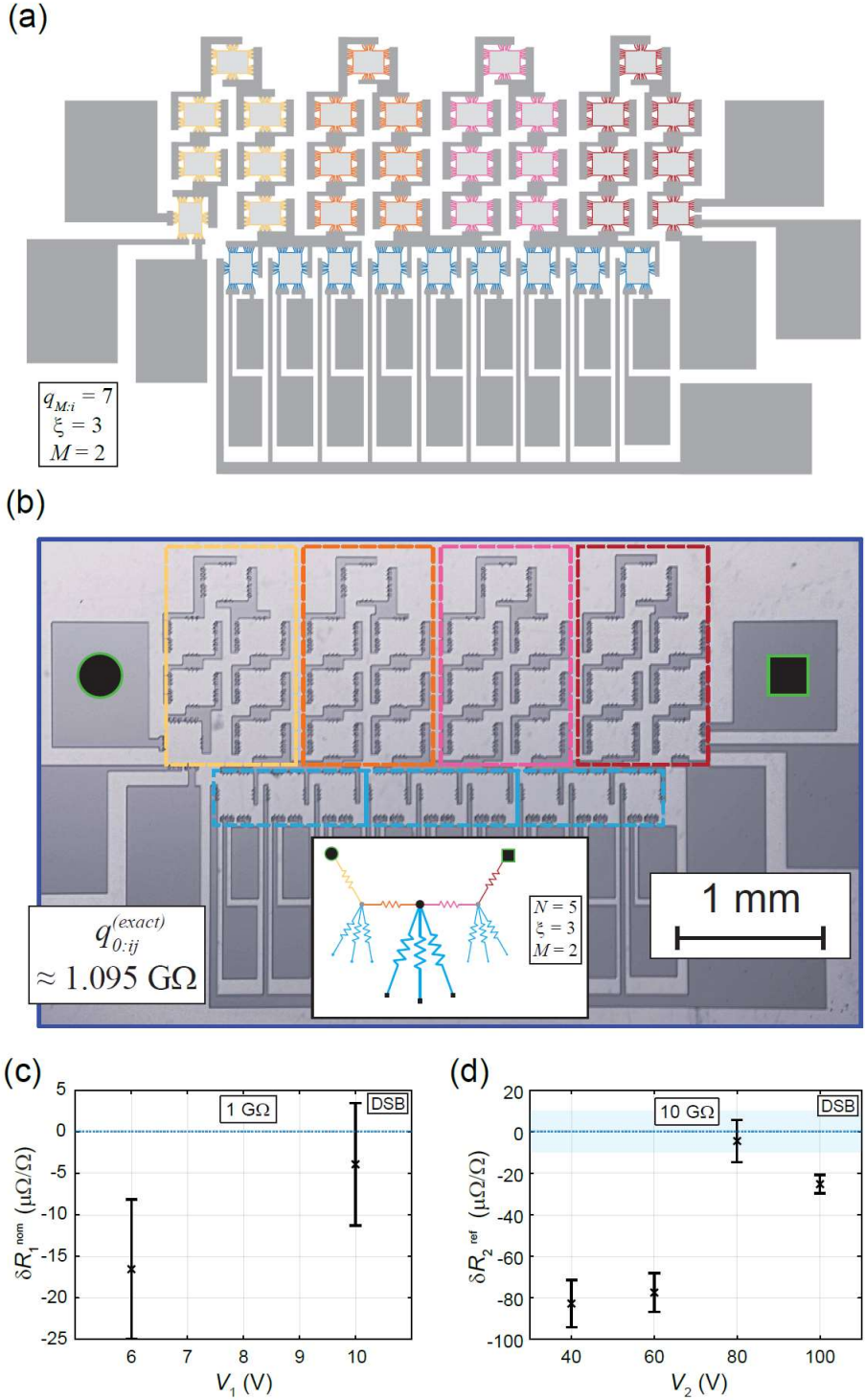

FIG. 2. (a) QHARS device design layout with corresponding parameters $M = 2$, $\xi = 3$, and $q_{M:i} = 7$. (b) Device post-fabrication optical image, overlaid by color-coded markers that match with those in the central inset showing a simplified network drawing presented in the style of Ref. [22]. (c) Precision measurement DSB data are shown to confirm the quantization of the QHARS device, with a standard resistor serving as the known quantity. The deviation from nominal $\delta R_1$ is centered around $q_{0:ij}^{(exact)} = 1.095\,069\ldots$ G$\Omega$. Error bars represent a 1$\sigma$ uncertainty. (d) More DSB data denote the utility in using QHARS devices for calibrations at 10 G$\Omega$. The highlighted blue is the 1$\sigma$ uncertainty on the value of the standard resistor and the data points are the determined value of the same resistor using the QHARS device, with error bars representing 1$\sigma$ uncertainty.

$$q_{m:\neg ij} = q_{m+1:ij}^{(exact)} + 2\left\{\xi + \sum_{x=m+1}^{M-1}\left[\frac{2}{q_{x:\neg ij}}\right]\right\}^{-1}$$

(5)

With all terms well-defined, one can apply this generalized approach to an example device design. The desired *effective* quantized resistance was 1 GΩ, and the QHARS design optimization yielded the device shown in Fig. 2 (a) and (b) (with the corresponding parameters $M = 2$, $\xi = 3$, and $q_{M:i} = 7$). Using Eq. 3, one finds that $q_{M:ij}^{(near-exact)} = 1.007\ 797\ ...$ GΩ. The LTspice simulation for the device matched the exact value found via Eq. 4: $q_{0:ij}^{(exact)} = 1.095\ 069\ ...$ GΩ. To support the simulation, precision measurements of the device were performed via dual source bridge (DSB), allowing for sensitivities on the order of μΩ/Ω [34–40]. See Supplemental Material for more context and information about calibrations [41]. These DSB data are shown in Fig. 2 (c) and (d) to confirm the quantization of the QHARS device using a 1 GΩ and 10 GΩ standard resistor, respectively.

In Fig. 2 (c), the deviation from nominal $\delta R_1$ is centered around $q_{0:ij}^{(exact)} = 1.095\ 069\ ...$ GΩ and the error bars represent a 1σ uncertainty. The 1 GΩ standard resistor served to assess and validate the QHARS device's functionality. The 10 GΩ DSB data in Fig. 2 (d) denote the utility in using QHARS devices for calibrations this level. The highlighted blue is the 1σ uncertainty on the value of the standard resistor calibrated via conventional traceability chain, whereas the data points are the determined value of the same resistor using the QHARS device as the known resistor, with error bars signifying a 1σ uncertainty. The deviation of the 10 GΩ resistor value from as determined by the QHARS device from the conventionally calibrated value is defined as $\delta R_2^{ref}$.

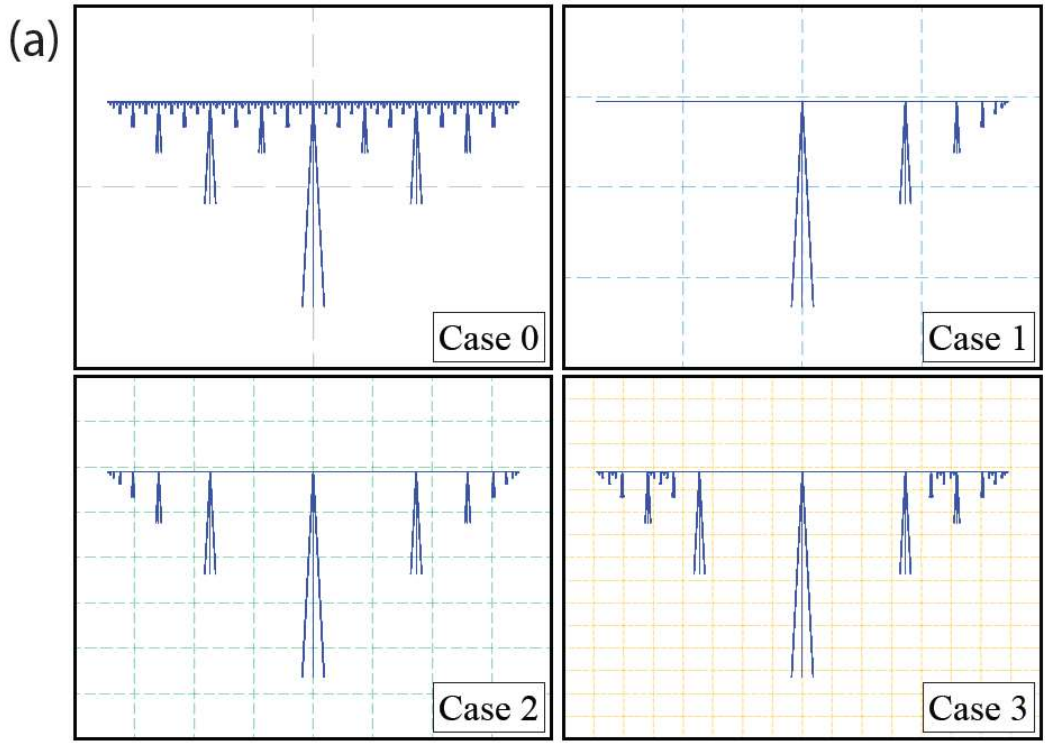

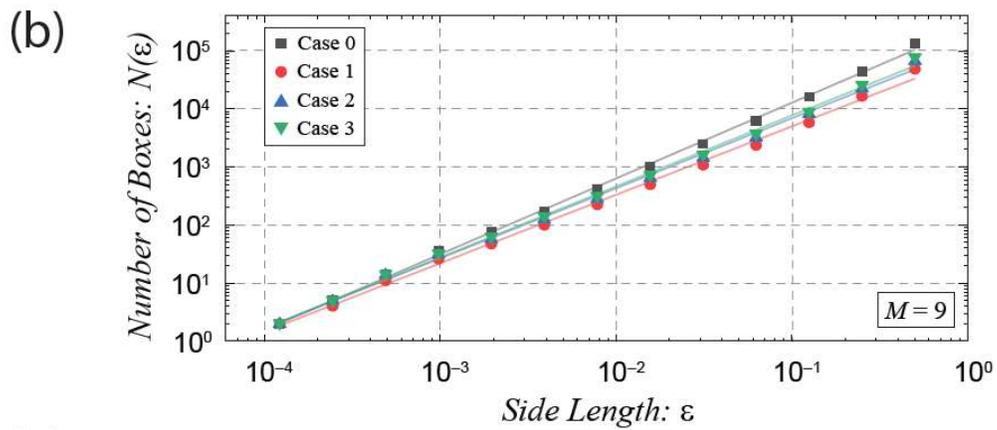

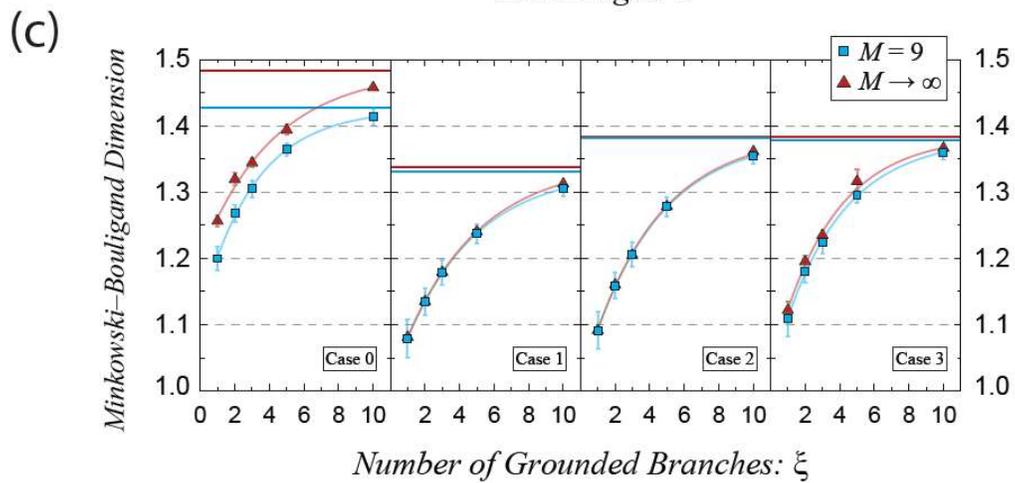

FIG. 3. (a) This illustration depicts the four distinct pseudofractals used for the Cases in the relevant calculations. Each panel is accompanied by an example overlay of a single box-counting computation, where the boxes containing the pseudofractal are tracked as a function of box size. (b) The prior computations are tabulated and plotted on a log–log plot, whose resulting curves may be fit to a line whose slope reveals the Minkowski–Bouligand dimension of the relevant pseudofractal. Example pseudofractals from the $M = 9$ iterations of all four Cases from (a) are plotted. (c) A summary of the MBDs for each pseudofractal are shown as a function of $\xi$, with two distinct curves for the same $M = 9$ iterations (in blue) and for the ideal case of the pseudofractal where the iterations approach infinity (in red).

Until now, the discussion has focused primarily on using full recursion to design a device in the neighborhood of a desired quantized resistance, with an option to manually calculate (or alternatively, simulate with LTspice) the exact, measurable value of the corresponding device design. One of the risks of using full recursion is that if any of the grounded branches fail, whether by poor quantization or contacting, the expected value of the device changes substantially. One way to both secure the benefit of rapidly approaching high *effective* quantized resistances and mitigate the aforementioned risk is to consider the implementation of other foundational pseudofractal designs (of which the full recursion, or Case 0, is just one). A cursory examination of potential pseudofractal designs for 1 GΩ reveals that more complex pseudofractals are correlated with a more rapid divergence from the neighborhood of desired resistance values in the event of a grounded branch failure (or deviation). To commence any correlation analysis between QHARS design pseudofractal complexity and some metric indicating the resilience or recoverable utility of a QHARS device, one must first mathematically derive some quantities for other pseudofractal designs, primarily by modifying the definition of $M$ to account for different kinds of recursions. To that end, three additional Cases were considered based on a previous work [26], each with a unique rule of iteration and thus a different definition of $M$. One should expect Eq. 2 to change based on the new definition.

Figure 3 (a) shows the three new Cases, with the first being a unidirectional partial recursion, the second being a bidirectional partial recursion, and the third being a hybrid of Cases 0 and 2. As Ref. [26] points out, the means of calculating the total number of elements in a QHARS device for Cases 1, 2, and 3, respectively, are as follows:

$$D_T(M, \xi, q_{M:ij}) = M\xi - \frac{(M+1)}{\xi} + \frac{1}{\xi}\left(\xi q_{M:ij}^{(approx)} + 1\right)^{2^{-M}} + \frac{1}{\xi}\sum_{x=1}^{M}\left(\xi q_{M:ij}^{(approx)} + 1\right)^{2^{-x}}$$

(6)

$$D_T(M, \xi, q_{M:ij}) = (2M-1)\xi + \frac{2}{\xi}\left(\xi q_{M:ij}^{(approx)} + 1\right)^{2^{-M}} - \frac{2}{\xi} + 2\sum_{x=2}^{M}\left[\frac{1}{\xi}\left(\xi q_{M:ij}^{(approx)} + 1\right)^{2^{-x}} - \frac{1}{\xi}\right]$$

(7)

$$D_T(M,\xi,q_{M:ij}) = 2\xi \sum_{x=1}^{M}\left[2^{\frac{x-H[(-1)^{x+1}]}{2}}-1\right] + \left(\frac{1}{\xi}(\xi q_{M:ij}^{(approx)}+1)^{2^{-M}}-\frac{1}{\xi}\right)*2^{\frac{M+H[(-1)^{M+1}]}{2}+H[(-1)^M]} + H\left[M-\frac{3}{2}\right]$$
$$*\sum_{x=1}^{M-1} 2^{\frac{x}{2}} * H[(-1)^x] * \left(\frac{1}{\xi}(\xi q_{M:ij}^{(approx)}+1)^{2^{-x}}-\frac{1}{\xi}\right)$$

(8)

These formulas will help extract the first variable, a metric of the divergence away from a neighborhood of desired resistances, but first, one must also define the second variable reflecting a pseudofractal complexity, since selecting certain design approaches has implications on recovering QHARS device utility in the event of a grounded branch failure. This second variable will be the Minkowski–Bouligand dimension (MBD), which characterizes a fractal or pseudofractal via a ratio of its change in detail to change in scale [26]. This quantity is calculated via the box-counting method of vector representations of the topological drawings shown in Fig. 3 (a), yielding the behavior in (b), whose linearity determines the value of the MBD. Example pseudofractals from the $M = 9$ iterations of all four Cases from Fig. 3 (a) are plotted, and Fig. 3 (c) shows a summary of the MBDs for each pseudofractal as a function of $\xi$, with two distinct curves for the same $M = 9$ iterations and for the ideal case of the pseudofractal where the iterations approach infinity.

Now that the MBDs are determined, one may focus on the metric of divergence, namely, an exponential growth factor for the total elements in a QHARS device for each of the Cases. This general way of correlating whether a design provides more localized neighborhoods of resistances involves inspecting the total device parameter from Eqs. 6 – 8 ($D_T$) [26]. Rapid divergence or rate of change of $D_T$ as $M \to \infty$ implies that minor unintended modifications (such as the failure of a single grounded Hall element) are more likely to result in larger changes in the output resistance of the QHARS device.

Each Case was examined by plotting $D_T$ on a semi-logarithmic scale in Fig. 4 (a). Calculations for the adjusted $D_T$ as a function of $M$ reflected pseudofractal designs near 1 GΩ. With each Case, the set of curves for $\xi = \{1, 2, 3, 5, \text{and } 10\}$ are logarithmically normalized such that the minimum curve values are at 10 for easier comparison [26]. The true total number of devices are in the Supplemental Material [41]. In the first panel of Fig. 4 (a), a dashed orange line is superimposed and horizontally translated to extract the exponential coefficient, essentially the rate of change of the adjusted $D_T$. Consistent with the results of Ref. [26], Cases 1 and 2 exhibit a low rate of change when compared with Cases 0 and 3, regardless of the value of $\xi$. This observation can be explained by the limiting behavior of Eqs. 6, 7, and 8; that is, when $M$ approaches large values in Cases 0 and 3, the behavior is exponential growth, and for Cases 1 and 2, it becomes linear growth since the terms containing exponential dependences on $M$ approach zero.

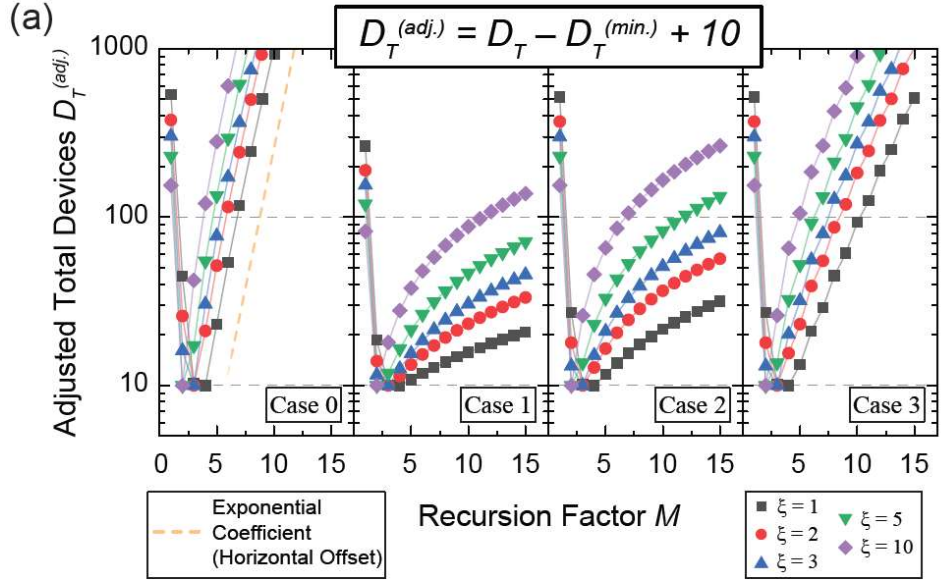

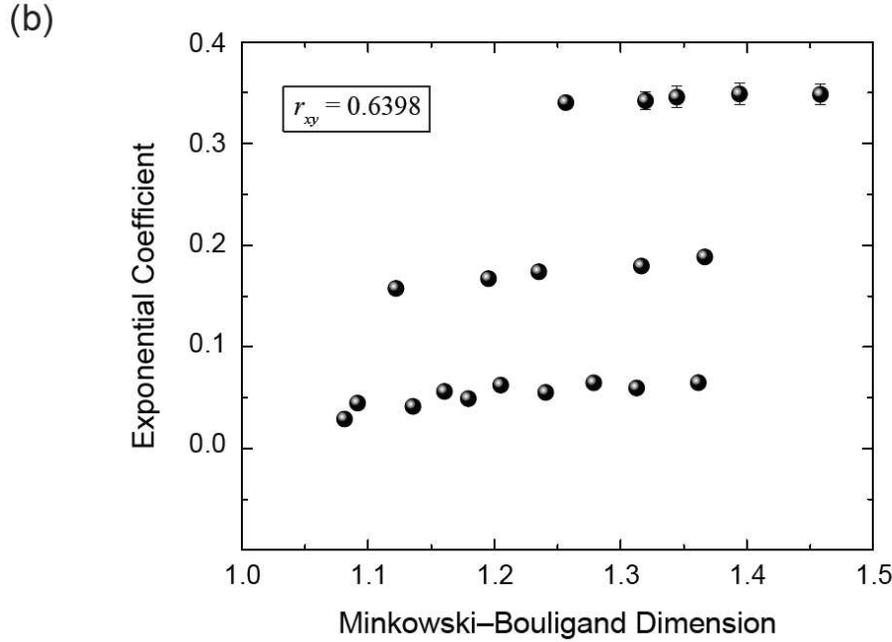

FIG. 4. (a) Calculations for the adjusted total devices as a function of recursion factor ($M$) are performed for each of the four Cases of pseudofractal design. The target resistance was 1 GΩ, and all plots are logarithmically normalized such that the minimum values are at 10 for easier comparison. The true total number of devices are in the Supplemental Material [41]. A horizontally offset exponential growth fit is imposed in the first panel. (b) The Pearson correlation coefficient is determined between the calculated Minkowski–Bouligand Dimension and the exponential coefficients in (a). The error bars indicate 1σ uncertainty.

With both variables now identified, the Pearson correlation coefficient was subsequently determined between the calculated MBD of a pseudofractal (one of four Cases) and the exponential coefficients in (a). The results from this correlation are in Fig. 4 (b) and the coefficient itself is valued at about 0.6398 ± 0.0077. This positive correlation suggests, as it did in Ref. [26], that for QHARS topologies with higher MBD, one is less likely to design QHARS devices that provide access to localized neighborhoods of resistance. Furthermore, it is a statement about pseudofractals exhibiting linear rather than exponential limiting behavior, a quality that QHARS device designers should utilize when seeking greater ease of access to a localized neighborhood of resistance values. Additional simulations for each of the four Cases are provided in the Supplemental Material to further illustrate the divergence of a device output in the event of a grounded branch failure [41].

This work refines a recently developed mathematical framework for minimizing the number of elements in a QHARS device and addresses slight discrepancies between desired and actual resistance values that arise from approximations within the initial framework. The analysis is supported by data from an approximately 1 GΩ QHARS device and prompts one to explore the benefits of other pseudofractal designs. By examining the MBD of these pseudofractals, one can correlate QHARS device complexity with device output resistance divergence from the desired neighborhood when subjected to grounded branch failure. This understanding provides significant advantages for designers of future QHARS device topologies.

## ACKNOWLEDGMENTS

The authors thank M. Munoz, F. Fei, and E. C. Benck for their assistance in the NIST internal review process, and A. R. Panna and Y. Yang for fruitful guidance. The authors declare no competing interests. Commercial equipment, instruments, and materials are identified in this paper in order to specify the experimental procedure adequately. Such identification is not intended to imply recommendation or endorsement by the National Institute of Standards and Technology or the United States government, nor is it intended to imply that the materials or equipment identified are necessarily the best available for the purpose. Work presented herein was performed, for a subset of the authors, as part of their official duties for the United States Government. Funding is hence appropriated by the United States Congress directly.

## DATA AVAILABILITY

Data that support the findings of this study are available from the corresponding author upon reasonable request.

# Supplementary Material: Influences of the Minkowski–Bouligand Dimension on Graphene-Based Quantum Hall Array Designs

D. S. Scaletta,[1] N. T. M. Tran,[2,3] M. Musso,[4] V. Ortiz Jimenez,[2] H. M. Hill,[2] D. G. Jarrett,[2] M. Ortolano,[4] C. A. Richter,[2] D. B. Newell,[2] and A. F. Rigosi[2,a]

[1]*Department of Physics, Mount San Jacinto College, Menifee, California 92584, USA*

[2]*Physical Measurement Laboratory, National Institute of Standards and Technology (NIST), Gaithersburg, Maryland 20899, USA*

[3]*Joint Quantum Institute, University of Maryland, College Park, Maryland 20742, USA*

[4]*Department of Electronics and Telecommunications, Politecnico di Torino, Torino 10129, Italy*

[b)] Author to whom correspondence should be addressed.  email: afr1@nist.gov

Table of Contents:

1. Additional context for Dual Source Bridges and calibrations
2. Logarithmic Normalization Values for Figure 4
3. Simulations of Grounded Branch Failures

# 1. Additional context for Dual Source Bridges and calibrations

For context, the DSB has been used as the primary method for the measurement of high resistance ranges (see main text Refs. [34-36]). The bridge is an adapted Wheatstone bridge, which is based on the automated high resistance measurement approach initially proposed by Henderson (see main text Ref. [37]). The DSB is configured with two voltage sources in the main ratio arms, substituting two corresponding resistors in the Wheatstone bridge and forming a voltage ratio bridge. When the bridge is balanced, the arm containing the unknown resistance value $R_x$ can be determined via $R_x = R_s \frac{V_x}{V_s}$, where $R_s$ is a standard resistor and $V_x$ and $V_s$ are the applied voltages across $R_x$ and $R_s$, respectively.

Each standard resistor was calibrated by a conventional traceability chain involving the use of a graphene-based quantized Hall resistance standard that is approximately 12.9 k$\Omega$, a cryogenic current comparator, and Hamon transfer standards to reach the two high resistance values (see main text Refs. [36, 38–40]).

# 2. Logarithmic Normalization Values for Figure 4

To better compare the behavior of the total device count in the main text, each curve was logarithmically normalized to its global minimum. This table summarizes those values of $D_T^{(min)}$, rounded to the nearest integer:

| 1 G$\Omega$ | $D_T^{(min)}$ | | | |
|---|---|---|---|---|
| Case | 0 | 1 | 2 | 3 |
| $\xi = 1$ | 31 | 302 | 49 | 49 |
| $\xi = 2$ | 28 | 215 | 36 | 36 |
| $\xi = 3$ | 30 | 178 | 34 | 34 |
| $\xi = 5$ | 34 | 144 | 34 | 34 |
| $\xi = 10$ | 41 | 114 | 41 | 41 |

## 3. Simulations of Grounded Branch Failures

To test, in practice, how each pseudofractal correlates with a divergent behavior of the QHARS output, several examples of grounded branch failures were simulated with LTspice.

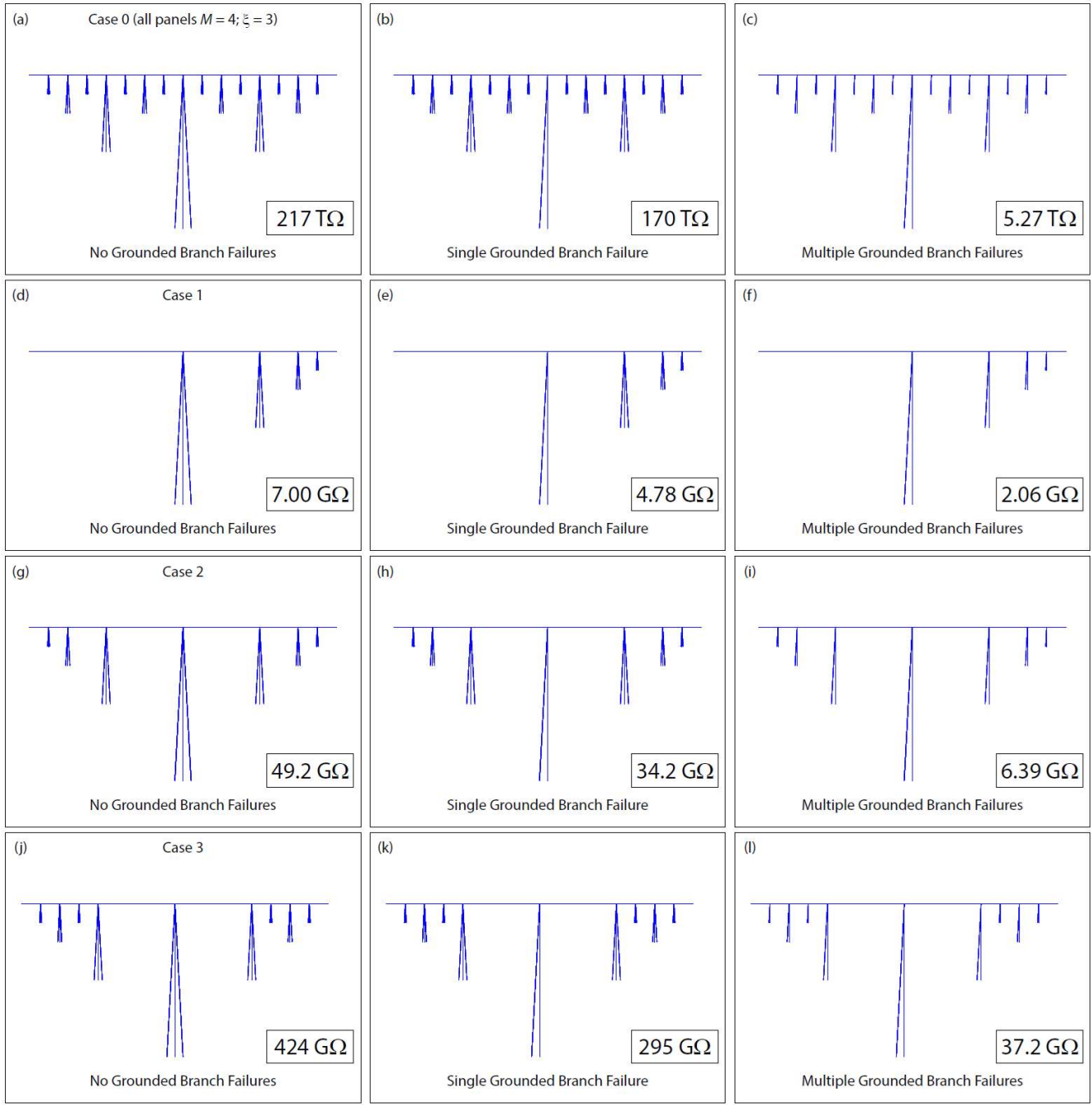

FIG 1-SM. Example simulations in LTspice to obtain grounded branch failure impacts on a QHARS device.